\documentclass[twocolumn,preprint]{aastex62}
\usepackage[T1]{fontenc}
\usepackage[english]{babel}
\usepackage{natbib}
\usepackage{ulem,xcolor}
\usepackage{amsmath, graphicx}
\usepackage{soul}
\def\rhessi{{\textit{RHESSI}}}
\def\IRIS{{\textit{IRIS}}}

\def\kw{{Konus-\textit{Wind}}}
\def\sdo{{\textit{SDO}}}

\def\mw{{microwave}}

\def\gx{{GX Simulator}}

\newcommand\gnout{\bgroup\markoverwith{\textcolor{red}{\rule[0.5ex]{2pt}{0.4pt}}}\ULon}

\begin{document}

\title{Data-constrained 3D modeling of a solar flare evolution: acceleration, transport, heating, and energy budget}

	\author[0000-0001-5557-2100]{Gregory D. Fleishman}
	\affil{Center For Solar-Terrestrial Research, New Jersey Institute of Technology, Newark, NJ 07102, USA
	}
	\affil{Leibniz-Institut für Sonnenphysik (KIS), Freiburg, 79104, Germany
	}
	\author[0000-0003-2846-2453]{Gelu M. Nita}
	\affil{Center For Solar-Terrestrial Research, New Jersey Institute of Technology, Newark, NJ 07102, USA
	}


\author[0000-0001-7856-084X]{Galina G. Motorina}
	\affil{Astronomical Institute of the Czech Academy of Sciences, 251 65 Ond\v{r}ejov, Czech Republic\\
Central  Astronomical Observatory at Pulkovo of Russian Academy of Sciences, St. Petersburg, 196140, Russia\\
Space Research Institute of Russian Academy of Sciences, Moscow, 117997, Russia}


\begin{abstract}
Solar flares are driven by release of the free magnetic energy and its conversion to other forms of energy---kinetic, thermal, and nonthermal. Quantification of partitions between these energy components and their evolution is needed to understand the solar flare phenomenon including nonthermal particle acceleration, transport, and escape and the thermal plasma heating and cooling. The challenge of remote sensing diagnostics is that the data are taken with finite spatial resolution and suffer from line-of-sight (LOS) ambiguity including cases when different flaring loops overlap and project one over the other. Here we address this challenge by devising a data-constrained evolving 3D model of a multi-loop SOL2014-02-16T064620 solar flare of GOES class C1.5. Specifically, we employed a 3D magnetic model validated earlier for a single time frame  and extended it to cover the entire flare evolution. For each time frame we adjusted {the} distributions of the thermal plasma and nonthermal electrons in the model, such as the observables synthesized from the model matched the observations. Once the evolving model has been validated this way, we computed and investigated the evolving energy components and other relevant parameters by integrating over the model volume. This approach removes the LOS ambiguity and permits to disentangle contributions from the overlapping loops. It reveals new facets of electron acceleration and transport, as well as heating and cooling the flare plasma in 3D. We find signatures of substantial direct heating of the flare plasma not associated with the energy loss of nonthermal electrons.







\end{abstract}

\keywords{Sun: Flares - Sun: X-rays, EUV, Radio emission}

\section{Introduction}
\label{S_Intro}

Solar flares observed as transient brightenings at various wavelengths,  occur sporadically on the Sun when free magnetic energy of an active region (AR) transforms suddenly to other forms of energy \citep{2011SSRv..159...19F}. This complex and yet poorly understood chain of physical transformations includes acceleration of charged particles, their transport, plasma heating, and macroscopic plasma motions \citep[e.g.,][]{2012ApJ...759...71E}. 

Understanding the relationships between the release of magnetic energy \citep{2020Sci...367..278F} and response of the solar atmosphere \citep{1985ApJ...289..414F,2013ApJ...778...68R,2014ApJ...781...43C,2015ApJ...808..177R,2015ApJ...813..133R,2017ApJ...848...39B,2018ApJ...856..149R,2021ApJ...912...25A,2022ApJ...928..190K} requires detailed information of the solar plasma evolution in the three-dimensional (3D) domain inscribing flare structural elements (e.g., flaring loops). This information cannot be obtained from observational data alone, as these data represent 2D projections of the 3D realm on the picture plane. Thus, observational data have to be complemented by realistic 3D modeling. However, a full 3D modeling that includes evolution of both magnetic field and associated evolution of thermal and nonthermal plasma components is not yet currently feasible \citep{2017ApJ...834...10R,2019NatAs...3..160C}. Therefore, various simplifying assumptions are often made, or only a portion of flare evolution is being modeled \citep[e.g.,][]{2018ApJ...859...17F}. For example, the 3D modeling can be performed in detail when the released magnetic energy is modest such as the magnetic structure in the flaring region does not change significantly during the course of the flare. 

Here, we analyze the SOL2014-02-16T064600 flare described by \citet{2021ApJ...913...97F}, which we find suitable for a detailed time-dependent modeling in 3D for the following reasons. This is a relatively weak, confined C1.5-class flare, which did not produce any noticeable plasma motion or magnetic reconfiguration until the late {phase of the} flare, which implies that there were no significant changes in the magnetic structure during the course of the flare. This {conclusion} is also consistent with no measurable variation of the AR magnetic energy in the series of nonlinear force-free field {(NLFFF)} reconstructions performed during 1-hour long interval covering the flare \citep{2021ApJ...913...97F}. Another advantage is that a 3D model of this flare, built based on a NLFFF reconstruction, offers a magnetic connectivity almost perfectly consistent with X-ray and EUV imaging data for at least one time frame 
\citep[e.g., fig. 13 in][]{2021ApJ...913...97F}. This means that the static  flaring flux tubes identified in the NLFFF 3D model by \citet{2021ApJ...913...97F} are likely suitable for quantitative data-constrained modeling evolution of the nonthermal electrons and thermal plasma.



\section{Overview of the SOL2014-02-16T064600 flare}

\citet{2021ApJ...913...97F} reported a detailed analysis of the available multi-wavelength data augmented by a context 3D model of one time frame of the flare. This flare displayed one episode of impulsive acceleration of electrons detected by microwave and X-ray emission, while three distinct episodes of plasma heating--both before and after the impulsive energy release.

\citet{2021ApJ...913...97F} revealed three distinct flux tubes with noticeably different properties involved in the SOL2014-02-16T064600 flare. 
Two of these flux tubes did not show any evidence of a significant nonthermal component at the peak time of the nonthermal emission. At this time, the nonthermal electrons were only detected in the largest and hottest loop (Flux Tube II; see below). Flux Tube II showed thermal-to-nonthermal behavior consistent with the Neupert effect \citep{1968ApJ...153L..59N}. However, all loops demonstrated noticeable pre-heating prior to the nonthermal electron component appearance. Thus, the energy deposition by the nonthermal electrons was insufficient to support the thermal response in this flare, favoring an additional plasma heating. 

\subsection{Observational data}

All data sources available for this event and their analysis were described in \citet{2021ApJ...913...97F}. Here, we use a subset of the data needed for modeling  the flare thermal and nonthermal components. Specifically, the composite \mw\  (Nobeyama Radio Polarimeter (NoRP, \citep{1979PRIAN..26..129T}), Radio Solar Telescope Network (RSTN, \citep{1981BAAS...13Q.553G}), and the Badary Broadband Microwave Spectropolarimeters (BBMS, \citep{2015SoPh..290..287Z}) and \kw\  X-ray  data\footnote{{\rhessi\ missed the flare impulsive phase due to its night, so it could not be used to quantify the nonthermal flare component.}} \citep{1995SSRv...71..265A, 2014Ge&Ae..54..943P} are used to constrain the nonthermal electron evolution, while \rhessi\ X-ray  \citep{2002SoPh..210....3L} and \sdo/AIA EUV data \citep{2012SoPh..275...17L} and data products are used to constrain the thermal plasma evolution.

\subsection{Master 3D model}
\citet{2021ApJ...913...97F} developed a 3D model based on the automated model production pipeline (AMPP) available in the GX Simulator distribution \citep{2023arXiv230100795N}. This model  {\citep[see animated fig. 11 in][]{2021ApJ...913...97F}} employs a standard NLFFF reconstruction produced from a preflare \sdo/HMI vector magnetograms at 06:34:12\,UT. The three flaring flux tubes were selected and populated by thermal plasma such as to match \rhessi\ spectral and imaging data, \IRIS\ \citep{2014iris} imaging in FeXXI line, and \sdo/AIA-derived emission measure (EM).  maps at 06:45:14\,UT. Then, Flux Tube II was populated with nonthermal electrons such as to match the \mw\ spectrum at the impulsive peak time of 06:44:41\,UT. The synthetic X-ray images generated from this model match the \rhessi\ imaging very closely, with only  a small spatial shift ($\lesssim2$"), which validates the model. The model shows that Flux Tubes I and II partly overlap; thus, their contributions to the thermal emission could not be properly separated based on the data analysis alone, which additionally calls for the time-dependent flare modeling performed here.


\section{Evolution of nonthermal electrons}

\subsection{Evolving distribution of nonthermal electrons in the flaring loops}
\label{S_F_evolution}

To recover evolution of nonthermal electrons in the flaring loops, we use a sequence of the microwave spectra available with 1\,s cadence \citep{2021ApJ...913...97F}. In the absence of the microwave imaging information, we have to make some reasonable assumptions about the locations where the nonthermal electrons are accelerated and reside. We assume: (i) the flaring flux tubes identified using 3D modeling of X-ray and EUV emissions {by \citet{2021ApJ...913...97F}} are the flux tubes where the nonthermal electrons reside and (ii) interaction between two flaring loops is responsible for {the} flaring process; thus, the initial acceleration happens where the loops intersect (or are in a close contact). 

{The microwave emission in solar flares is typically a combination of gyrosynchrotron and free-free contributions \citep{1998ARA&A..36..131B}. The theory of gyrosynchrotron emission generated by an electron distribution over the energy and pitch-angle \citep{Melrose_1968,1969ApJ...158..753R} offers exact analytical but cumbersome expressions for the emissivity and the absorption coefficient, whose straightforward numerical implementations result in very slow computer codes. The radiation transfer codes that \gx\ relies on  \citep{2023arXiv230100795N}  employ fast gyrosynchrotron codes \citep{2010ApJ...721.1127F} that includes both gyrosynchrotron and free-free components. Here, we use the fastest, continuous mode of the fast codes \citep[see][for the code nomenclature]{2010ApJ...721.1127F}.}

\gx\ permits rather sophisticated analytical shapes of the nonthermal electron distribution function over the energy, pitch-angle, and space. To model emission from different time frames of the given event requires varying some parameters of the distribution function, which brings in the dependence of time: 

\begin{equation}
\label{Eq_disfun}
   F(E,\mu,s,x,y,t)=F_E(E,t)F_\mu(\mu,t)F_s(s,t)F_r(x,y,s,t),
\end{equation}
where
\begin{equation}
\label{Eq_disfun_s}
    F_s(s,t)=\exp\left(-(q_0(t)(s+q_2(t))\right)^2-(q_1(t)(s+q_2(t)))^4),$$$$
     F_r(x,y,s,t)=\exp\left(-\left(\frac{p_0(t)x}{a(s)}\right)^2-\left(\frac{p_0(t)y}{b(s)}\right)^2\right),
\end{equation}
$s$ is the coordinate along the flux tube axes normalized by its length, such as $s_{\max}-s_{\min}=1$ and $s=0$ at the loop apex (where the magnetic field is minimal), $q_0$, $q_1$, $q_2$, and $p_0$ are free parameters of the distribution, $a(s)$ and $b(s)$ are the semi-axes of the elliptical transverse cross-section of the flux tube; they are computed based on the magnetic flux conservation and, thus, they are different along the loop axes. In this study we adopt a circular cross-section $a(s)=b(s)$.
The energy and angular distributions can be selected from a list of predefined analytical functions. Here we adopt a single power-law distribution over the energy $E$ and an isotropic angular distribution:

\begin{equation}
\label{Eq_disfun_Emu}
    F_E(E,t)=n_{\rm b}\frac{(\delta-1)E_{\min}^{\delta-1}}{E^\delta}; \ {\rm at}\  E_{\min}<E<E_{\max}; $$$$ F_\mu(\mu,t)=1/2,
\end{equation}
where $n_b$ is the number density of the nonthermal electrons between the minimum  $E_{\min}$ and the maximum $E_{\max}$ energies of the distribution (which is assumed to be zero outside this range) at the spatial location where $F_s=F_r=1$, and $\delta$ is the spectral index of the energy distribution. 

\begin{figure*}\centering
\includegraphics[width=0.8\linewidth]
{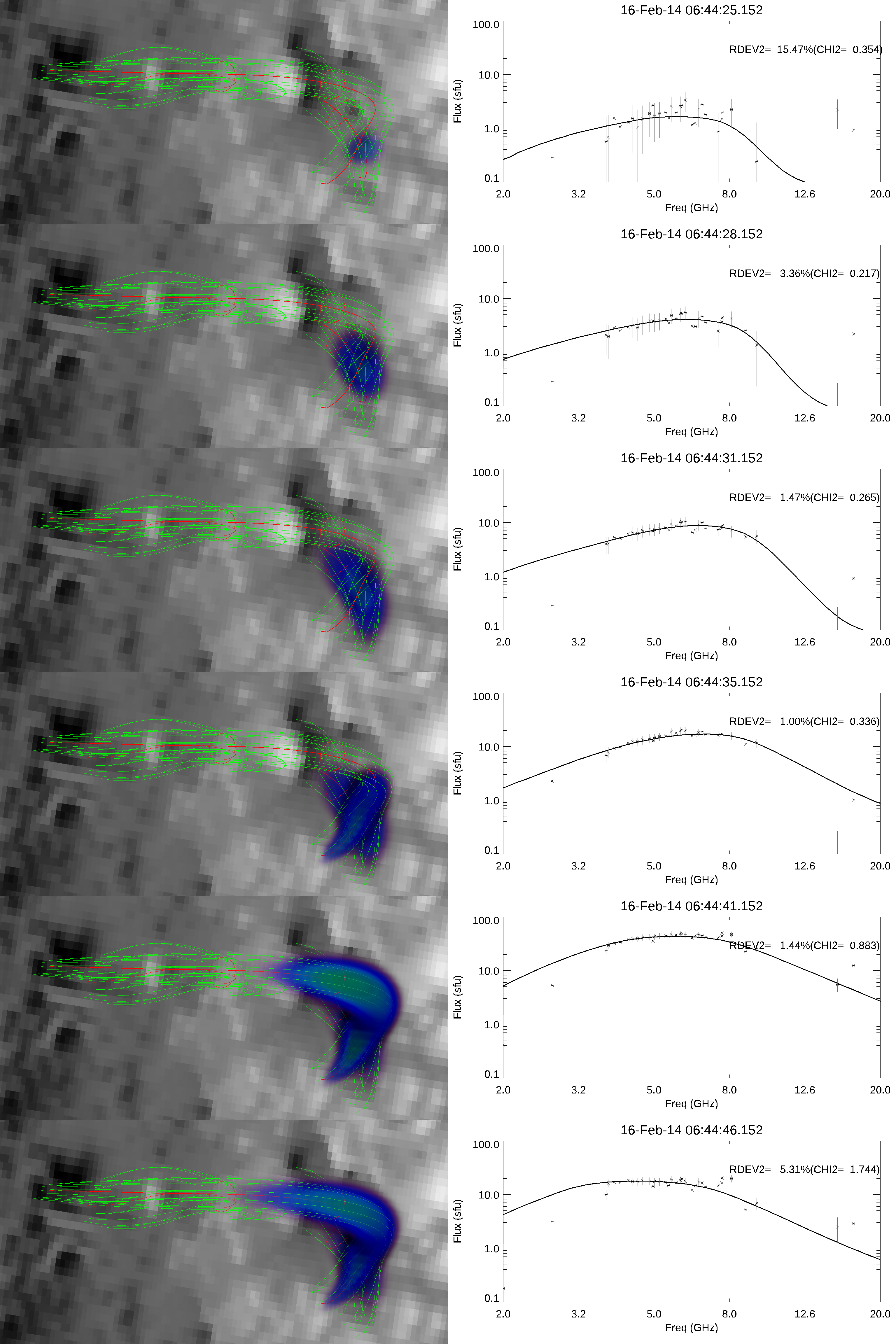}
\caption{
{Six representative snapshots {(left column)} showing evolution of nonthermal electrons in Flux Tubes I and II needed to reproduce evolution of the impulsive \mw\ emission {(right column). The background images show (the same) LOS HMI magnetogram; red lines show the selected central field lines of the model flaring flux tubes, while the green/blue volume filling visualize the number density of the nonthermal electrons needed to reproduce the observed \mw\ spectrum. The observed spectra are shown by the symbols with error bars at 1$\sigma$ level; the solid lines show the spectrum synthesized from the model.    An animation is included with this figure. The video shows a full sequence of the time frames with the perspective and top views of the evolving 3D model, evolving synthetic and observed spectra, and their residuals. The video
duration is 18 s.}}
\label{f_nonth_el_evol}
}
\end{figure*}

Finding the model distributions of the nonthermal electrons in the flaring loops is performed by trials and errors \citep{2018ApJ...859...17F} following established dependences of the \mw\ spectrum on the source parameters \citep[see, e.g., supplemenary video 3 in][]{2022Natur.606..674F}. As we have mentioned, given the lack of the \mw\ imaging data, we require that (i) the acceleration starts where two of the flaring loops may interact and (ii) the parameters show a reasonably smooth behavior in time. In addition, we attempted to keep the number of varying physical parameters minimal for each flux tube. We found that to fulfill requirement (i) we have to originally place the nonthermal electrons in a more compact Flux Tube I---in a location, where the axes of Flux Tubes I and II intersect in projection, see Figure\,\ref{f_nonth_el_evol}, top. The cloud of the nonthermal electrons appears as a very compact blob, which then extends in the volume. In about 9 seconds after the acceleration has started in Flux Tube I, the nonthermal electrons appear in Flux Tube II in much larger numbers than in Flux Tube I (see Figure\,\ref{f_nonth_el_evol}); thus, the nonthermal \mw\ emission at the impulsive peak is dominated by contribution from Flux Tube II \citep[cf.][]{2021ApJ...913...97F}.

\begin{figure}\centering
\includegraphics[width=1\linewidth]{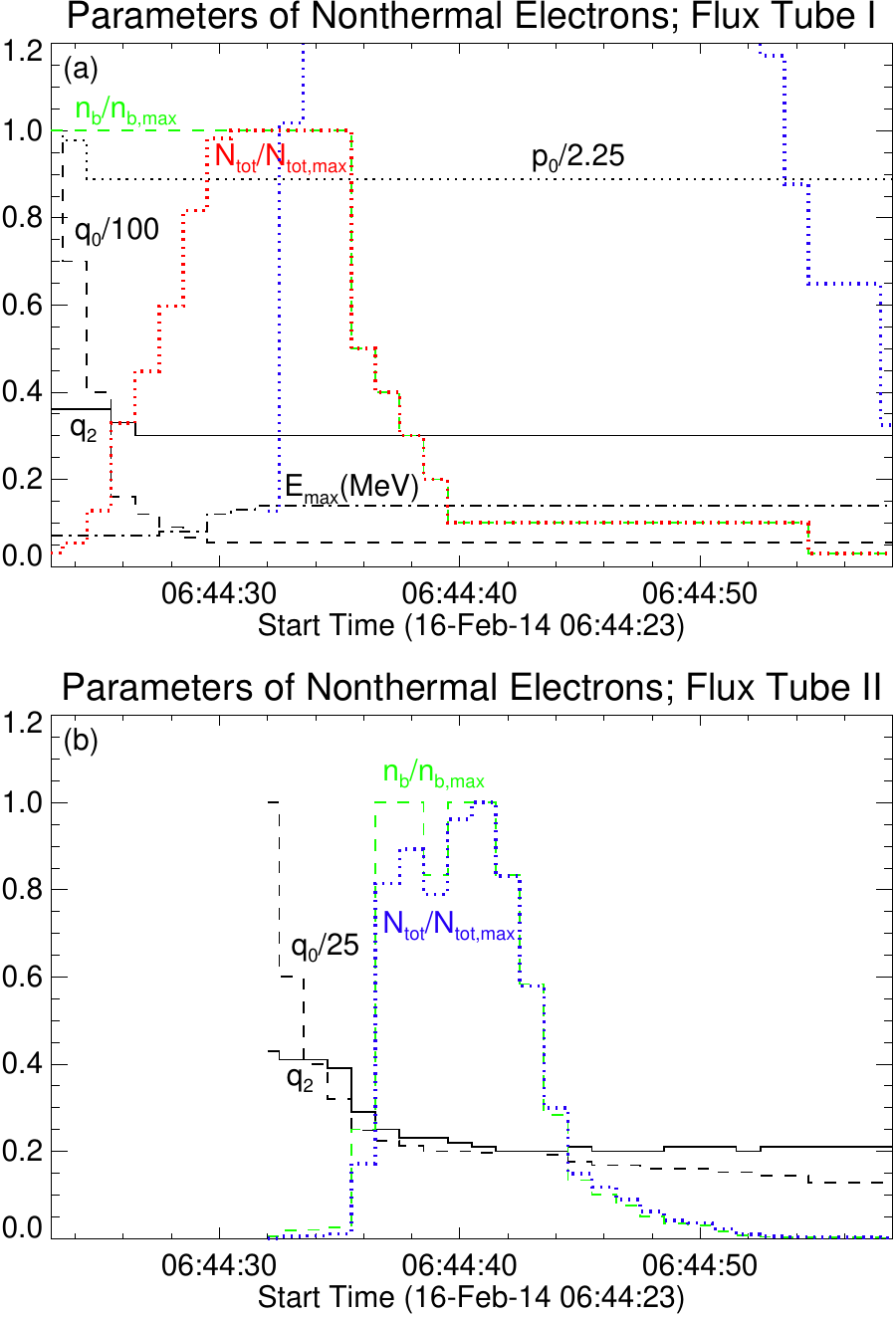}
\caption{
{Reconstructed evolution of the parameters of Flux Tube I (a) and II (b) defining the distribution functions of the nonthermal electrons; all curves are labeled in the panels. In Flux Tube I $n_{b, \max}(I)=10^8$\,cm$^{-3}$ and $N_{\rm tot, {\max}}(I)=1.11\times10^{33}$, while in  Tube II $n_{b, \max}(II)=1.2\times10^{10}$\,cm$^{-3}$ and $N_{\rm tot, {\max}}(II)=2.9\times10^{35}$. The blue dotted line in panel (a) displays the ratio $N_{\rm tot}(II)/N_{\rm tot, {\max}}(I)$ to be directly compared with the red line showing $N_{\rm tot}(I)/N_{\rm tot, {\max}}(I)$.}
\label{f_parms_nthm}
}
\end{figure}

After several runs of trials and errors, we determined that some parameters can be kept constant and the same in both flux tubes, namely, $\delta=3.9$ and $E_{\min}=10$\,keV. The geometry of the flux tubes was also kept unchanged with the reference transverse radius $a(0)=2.5$ grid points for Flux Tube I and $a(0)=2$ grid points for Flux Tube II (1 grid point $\simeq 1.04$'').
The set of varying parameters for Flux Tube I includes: $n_b$, $E_{\max}$, $q_0$, $q_2$, and $p_0$, although the last one, $p_0$, which controls the transverse distribution of the nonthermal electrons, was required to be slightly varied only during two first time frames; see Figure\,\ref{f_parms_nthm}a. For Flux Tube II, it was sufficient to  vary $n_b$, $q_0$, and $q_2$ only; see Figure\,\ref{f_parms_nthm}b. $E_{max}$ is not well constrained by data in Flux Tube II, so we adopted $E_{\max}=2$\,MeV. In Flux Tube I we adopted $q_1=0$, while in Flux Tube II, $q_1(t)=q_0(t)$; otherwise, the longitudinal distribution in Flux Tube II appeared too extended to fit the \mw\ spectra.

Let us consider the key features of the found solution displayed in Figure\,\ref{f_parms_nthm}. A highly remarkable property of the electron acceleration in Flux Tube I is that the number density of the nonthermal electron component is rather large, $n_b(I)\approx10^8$\,cm$^{-3}$, from the very start of the impulsive nonthermal emission. Moreover, this value does not grow in time: it stays constant at this level for 12 seconds and then rapidly falls down. Meanwhile, the \mw\ flux increases during this time interval due to an increase of the total number of the nonthermal electrons by a factor of 30---from $\approx3.5\times10^{31}$ to $\approx1.1\times10^{33}$ electrons. This increase is solely associated with the increase of the volume, where these nonthermal electrons reside, by the same factor from $\approx3.5\times10^{23}$\,cm$^3$ to   $\approx1.1\times10^{25}$\,cm$^3$. Another interesting feature required by the data is an increase of the maximum electron energy from 70 to 140\,keV during the first 10 seconds of the impulsive \mw\ emission, which can be interpreted as a time-dependent gain of the energy by nonthermal electrons---the very process of the particle acceleration.

A detectable amount of the nonthermal electrons appears in Flux Tube II 9 seconds later than in Flux Tube I, but it increases quickly such as, just in a couple of seconds, the total number of the nonthermal electrons in Flux Tube II gets larger than in Flux Tube I; see the intersection between the blue and red lines in Figure\,\ref{f_parms_nthm}a. Both the peak number density and the total number of the nonthermal electrons increase quickly during 5-6 seconds; then the number density stays almost constant during the next 5 seconds (while the total number still increases due to the volume increase), and then decreases during the remaining 17 seconds of the impulsive peak. Note, that the peak values of the number density are rather large, $n_{b, \max}(II)=1.2\times10^{10}$\,cm$^{-3}$. This corresponds to $N_{\rm tot, {\max}}(II)=2.9\times10^{35}$ nonthermal electrons in the effective volume $V_{eff}(II)=N_{\rm tot, {\max}}(II)/n_{b, \max}(II)\approx2.5\times10^{25}$\,cm$^3$. This modeling shows that the \mw\ data do not require any spectral evolution of the nonthermal electron population, but require a prominent evolution in space.

\subsection{Decoupling electron injection from trapping}
\label{S_trapping}

To quantify {the} energy deposition of the nonthermal electrons to the flaring loops we need to know their injection rate, rather than the evolving distribution function itself. \citet{2021ApJ...913...97F} roughly estimated this rate by noting that the \mw\ light curves are delayed relative to the \kw\ X-ray light curve by about 1 second, which was interpreted as the trapping/escape time $\tau$ from Flux Tube II. Here we attempt to perform a more accurate decoupling of the injection and trapping functions following the treatment of the trapping proposed by \citet{Fl_2005n}:

\begin{equation}
\label{Eq_transport}
 F(E,\mu,s,t)=\int_{-\infty}^t \exp\left(
 -\frac{t-t'}{\tau(E,\mu,s)}\right) G(E,\mu,s,t')dt'
\end{equation}
where $G(E,\mu,s,t')$ and $F(E,\mu,s,t)$ are the injection and
distribution functions of the fast electrons at the flaring loop axes
vs energy $E$, the cosine of pitch angle $\mu$, position along the
loop $s$, and time $t$. The physical meaning of this distribution function is the same as the one introduced by Eqn.\,(\ref{Eq_disfun}) at the loop axes---where $F_r=1$. In the general case, the phenomenological life-time
parameter $\tau$ can depend on the energy, position, and 
pitch-angle \citep{Fl_2005n}. Here we employ a simplified treatment guided by already specified (see the previous subsection) properties of the distribution function $F$---it is isotropic and does not show a noticeable spectral evolution. Thus, we adopt that $\tau$ does not depend of $E$ and $\mu$, but can depend on $s$. 
In our tests we found that a solution exists for a constant $\tau=1$\,s; thus, we ignore a possible dependence on $s$.

\begin{figure*}
\centering
\includegraphics
{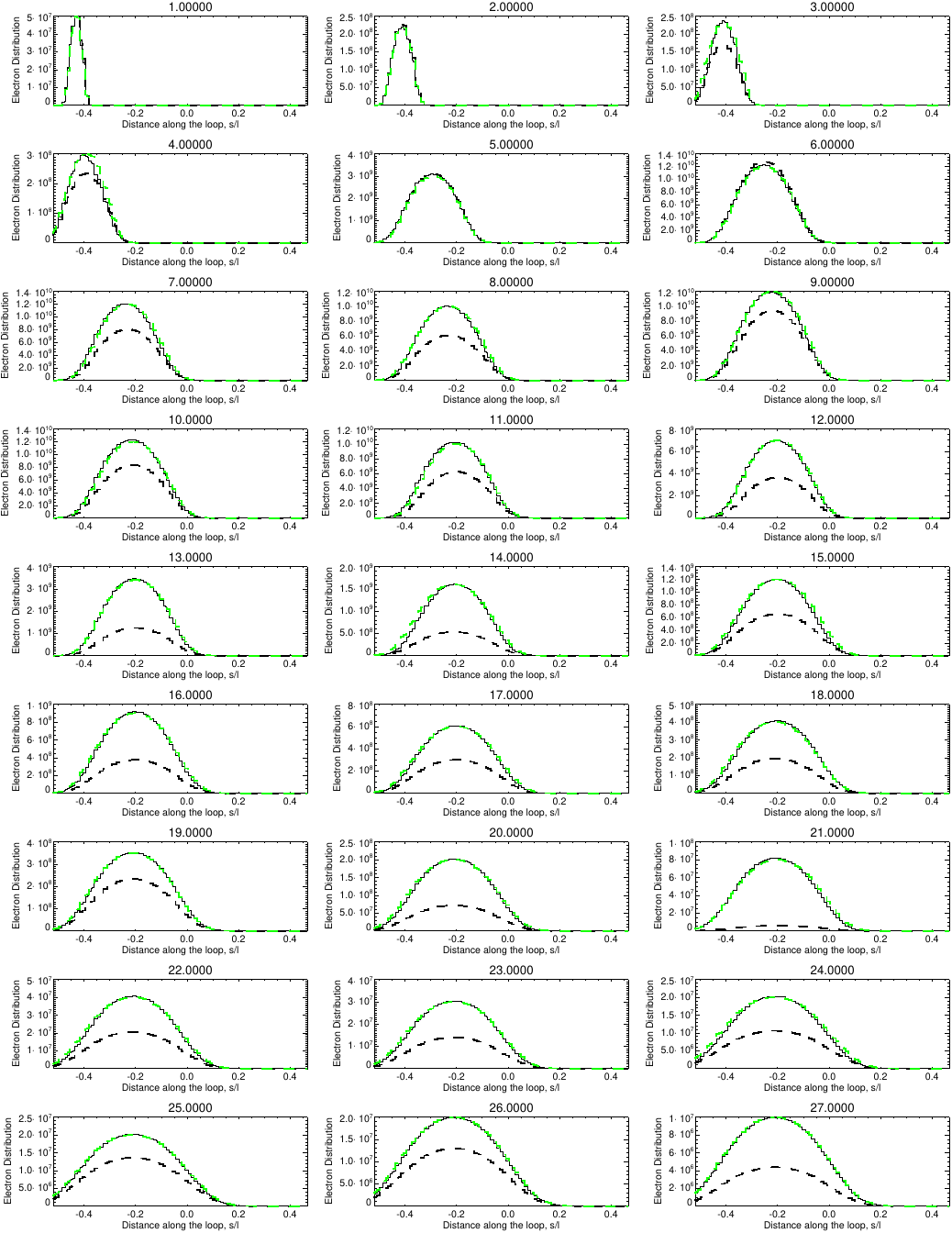}
\caption{
{Evolution of the longitudinal distribution functions of the nonthermal electrons obtained from Eqn.(\ref{Eq_transport}), solid black, and defined in Section\,\ref{S_F_evolution}, solid green. The corresponding injection profile is shown by the dashed solid lines. Each panel is scaled separately.}
\label{f_nonth_el_evol_s_2max}
}
\end{figure*}

\begin{figure*}\centering
\includegraphics
{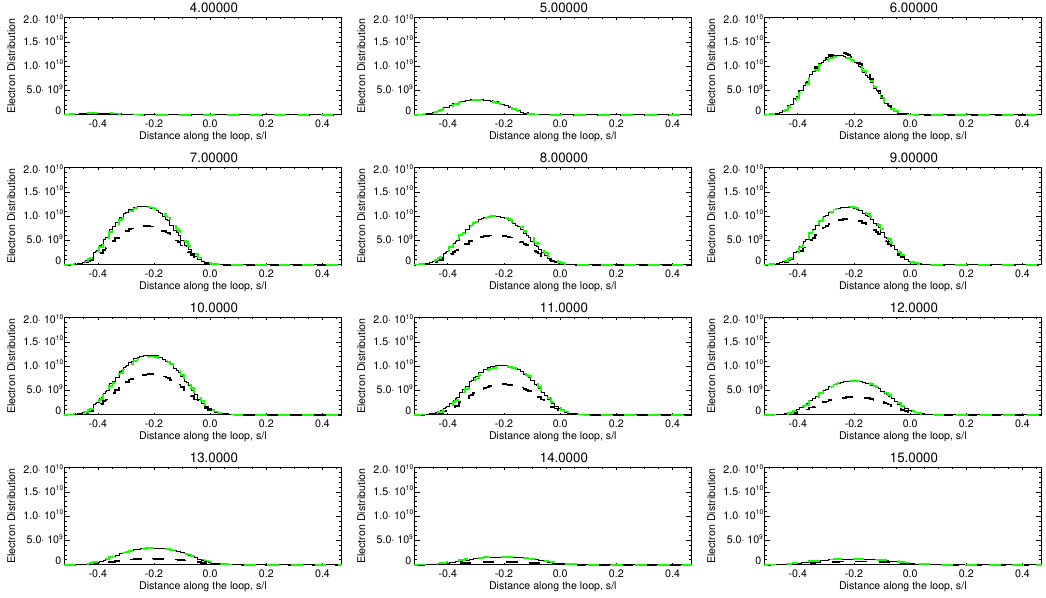}
\caption{
{A subset of panels from Fig.\,\ref{f_nonth_el_evol_s_2max}, but in the same scale normalized to the absolute maximum.}
\label{f_nonth_el_evol_s_2scale}
}
\end{figure*}

Given that the trapping time is about 1\,s, the distribution function at a given time $t$ is mainly defined by the injection function during the preceding one second---between $t-1$ and $t$. Earlier times give proportionally smaller contribution to $F$. Thus, we adopted the spatial shape of the injection function $G(E,s)$ during one second between $t-1$ and $t$ to be the same as the shape of the distribution function $F(E,s,t)$ found in Section\,\ref{S_F_evolution}, but with a generally different normalization. Given the available time cadence of the \mw\ data, we adopted 1\,s time bins for the injection and distribution functions. We considered sequentially the 27 time frames, where the distribution function in Flux Tube II was available, and defined the time-dependent normalization of the injection function $G$ by means of Eqn.\,(\ref{Eq_transport}), frame by frame. The solution is displayed by the solid lines in Figure\,\ref{f_nonth_el_evol_s_2max} along with the identified injection profile (dashed lines) and the solution identified in Section\,\ref{S_F_evolution} directly from comparison with the data (green lines). The success of the the solution is confirmed by almost perfect match between the solid black and green lines. Figure\,\ref{f_nonth_el_evol_s_2scale} displays a subset of the same panels during the peak of the emission, but now plotted in the same scale. One can see that the injection and distribution functions match each other\footnote{Although the injection rate is measured is cm$^{-3}$s$^{-1}$, while the number density is in cm$^{-3}$, we use the injection rate integrated during 1 s, thus, it has the same units as the number density, which permits direct comparison of these values.} at the raise phase. In contrast, in the decay phase the injection function is always smaller than the distribution function because here the contribution of previously injected electrons is not negligible.



\begin{figure}\centering
\includegraphics[width=1\linewidth]{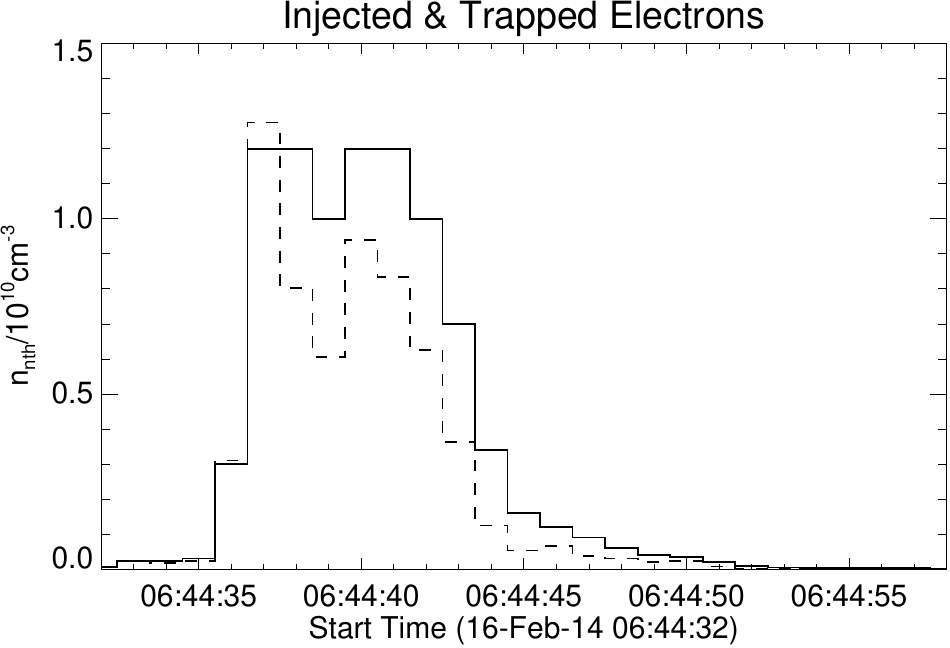}
\caption{
{Comparison of the spatial peak of the injection function (dashed line; cm$^{-3}$s$^{-1}$) and spatial peak of the nonthermal electron number density (solid line).}
\label{f_Inj_and_trapped}
}
\end{figure}

Figure\,\ref{f_Inj_and_trapped} compares the time profiles of the peak number density of the nonthermal electrons with the peak value of the injection function. Although the trapping time is relatively small, its effect is well seen, especially---at the decay phase, where the number density of nonthermal electrons is always larger, due to the trapping, than the number density injected during one previous second. 

\begin{figure}\centering
\includegraphics[width=1\linewidth]{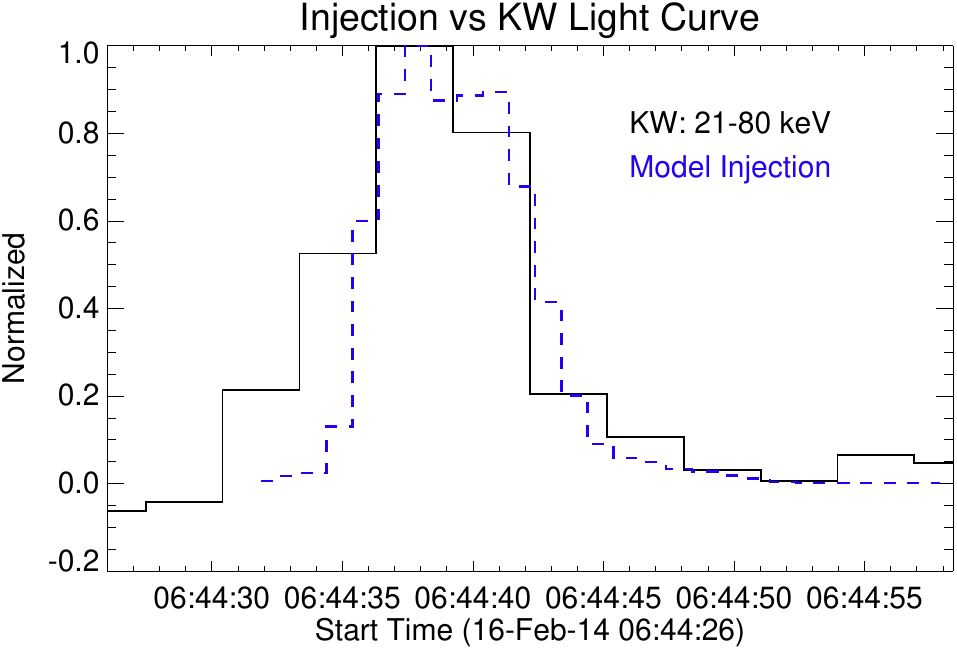}
\caption{
{Comparison between the injection function smoothed over 3\,s to match the \kw\ cadence (dashed blue) and the \kw\ HXR light curve (solid black). }
\label{f_Inj_vs_KW}
}
\end{figure}

Figure\,\ref{f_Inj_vs_KW} compares the \kw\ {hard X-ray} (HXR) light curve at 21--80\,keV obtained in waiting mode with the cadence of 2.944\,s and the model injection time profile smoothed over three seconds to match the \kw\ cadence. The two curves show a remarkable similarity, which further confirms the validity of the identified injection function.

\newpage

\section{Thermal plasma}

Properties and evolution of the thermal plasma in the 2014-Feb-16 flare are quantified by several data sets. \sdo/AIA data provide evolving maps of the EM derived from the reconstructed distributions of the differential EM (DEM), DEM-weighted temperature ($T$), and the thermal energy $W$ computed based on these $EM$ and $T$ {data reported by \citet{2021ApJ...913...97F}, where standard coronal ion abundances were adopted.} The methodology of {deriving} $EM$ and $T$ maps was described in detail by \citet{2020ApJ...890...75M}.  \rhessi\ provides evolving X-ray spectra and images. The spectral data yield spectral fit parameters within an adopted spectral model. Given that the \rhessi\ observations started roughly half a minute after the impulsive flare phase, here we employ a purely thermal {(using CHIANTI database with standard coronal abundances)} two-component spectral model with $T_1$ and $EM_1$ for a cooler component and $T_2$ and $EM_2$ for a hotter component. \IRIS\ provides some limited information on the 10\,MK plasma component based on the FeXXI line, but we do not use these data here, because this line was rather weak and not suitable for quantitative analysis \citep{2021ApJ...913...97F}.

\begin{figure}\centering
\includegraphics[width=1\linewidth]{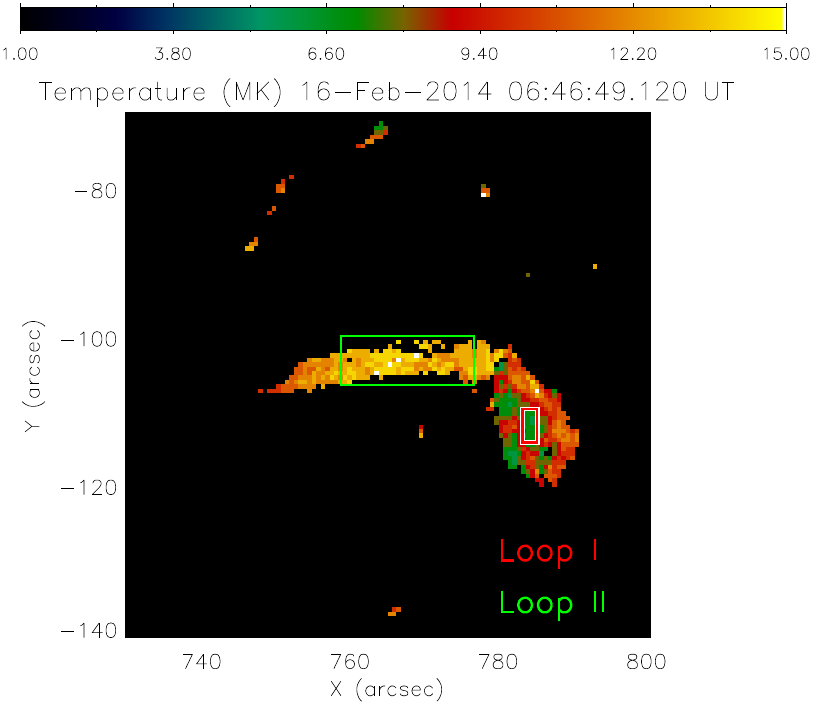}
\caption{
{Plasma temperature distribution obtained for time interval 06:46:37–06:46:49\,UT inferred from the DEM map \citep{2021ApJ...913...97F}. Red and green boxes, respectively, show the areas over which the mean temperatures of Flux Tubes I and II were evaluated. Only ``valid'' pixels, where the thermal energy estimated in \citet{2021ApJ...913...97F} is above 20\% of its peak value are taken into account.}
\label{f_temp_map}
}
\end{figure}

\begin{figure}\centering
\includegraphics[width=1\linewidth]{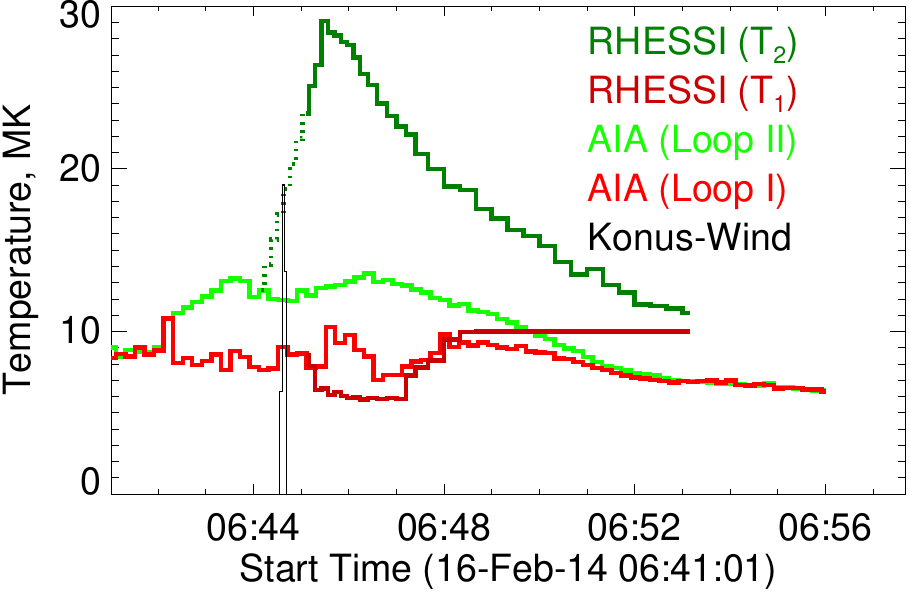}
\caption{
{Evolution of the plasma temperature inferred from the data. Dark green line is the larger temperature ($T_2$) and the dark red one is the smaller one ($T_1$) from the two-thermal-component \rhessi\ spectral fit reported by \citet{2021ApJ...913...97F}. \rhessi\ record started at 06:45:06\,UT; the dotted dark green line shows implied temperatures obtained from linear extrapolation of \rhessi\ data to the earlier time. The red and green lines show mean temperatures over valid pixels in the AIA DEM-weighted temperature maps inside two boxes covering portions of Flux Tubes I and II (shown in Fig.\,\ref{f_temp_map}). For the reference, the thin black line shows a conveniently scaled impulsive HXR emission  peak recorded by \kw .}
\label{f_temp_from_data}
}
\end{figure}

Figure\,\ref{f_temp_from_data} combines the flaring plasma temperatures derived from the data directly. The green and red lines display mean temperatures computed from two boxes covering two areas that project onto flaring loops II (hotter) and I (cooler), respectively, while the dark red and dark green lines display temperatures $T_1$ and $T_2$ obtained from the \rhessi\ spectral fit. For the reference, the solid black line shows the impulsive HXR emission. 

\begin{figure}\centering
\includegraphics[width=1\linewidth]{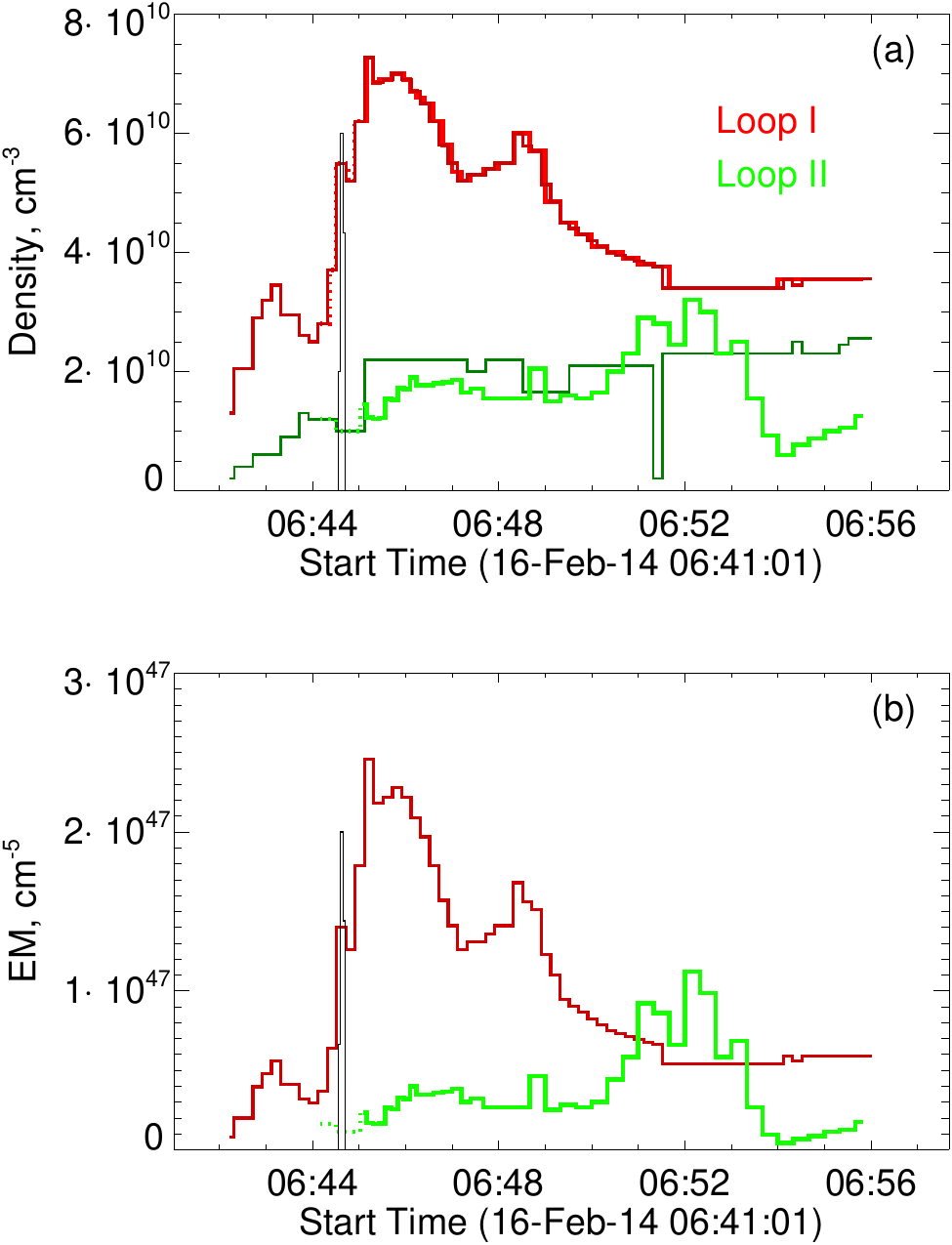}
\caption{
{Evolution of the spatial peak thermal density (a) and volume-integrated emission measure (b) in Flux Tubes I (red colors) and II (green colors). (a) Dark thin lines show the number densities inferred form the model adjusting to AIA EM maps. Green thick line shows the number density in Flux Tube II obtained from adjusting X-ray \rhessi\ spectra. Red thick line shows the number density values used to match the X-ray spectra--these values are close to those shown in dark red, but shifted to the closest \rhessi\ time stamp (as the \rhessi\ and AIA cadences do not match each other). (b) volume-integrated EM obtained from AIA EM map match for Flux Tube I and from the \rhessi\ spectral match for Flux Tube II. For the reference, the thin black line shows a conveniently scaled impulsive HXR emission  peak recorded by \kw .}
\label{f_model_n_EM}
}
\end{figure}

\begin{figure}\centering
\includegraphics[width=1\linewidth]{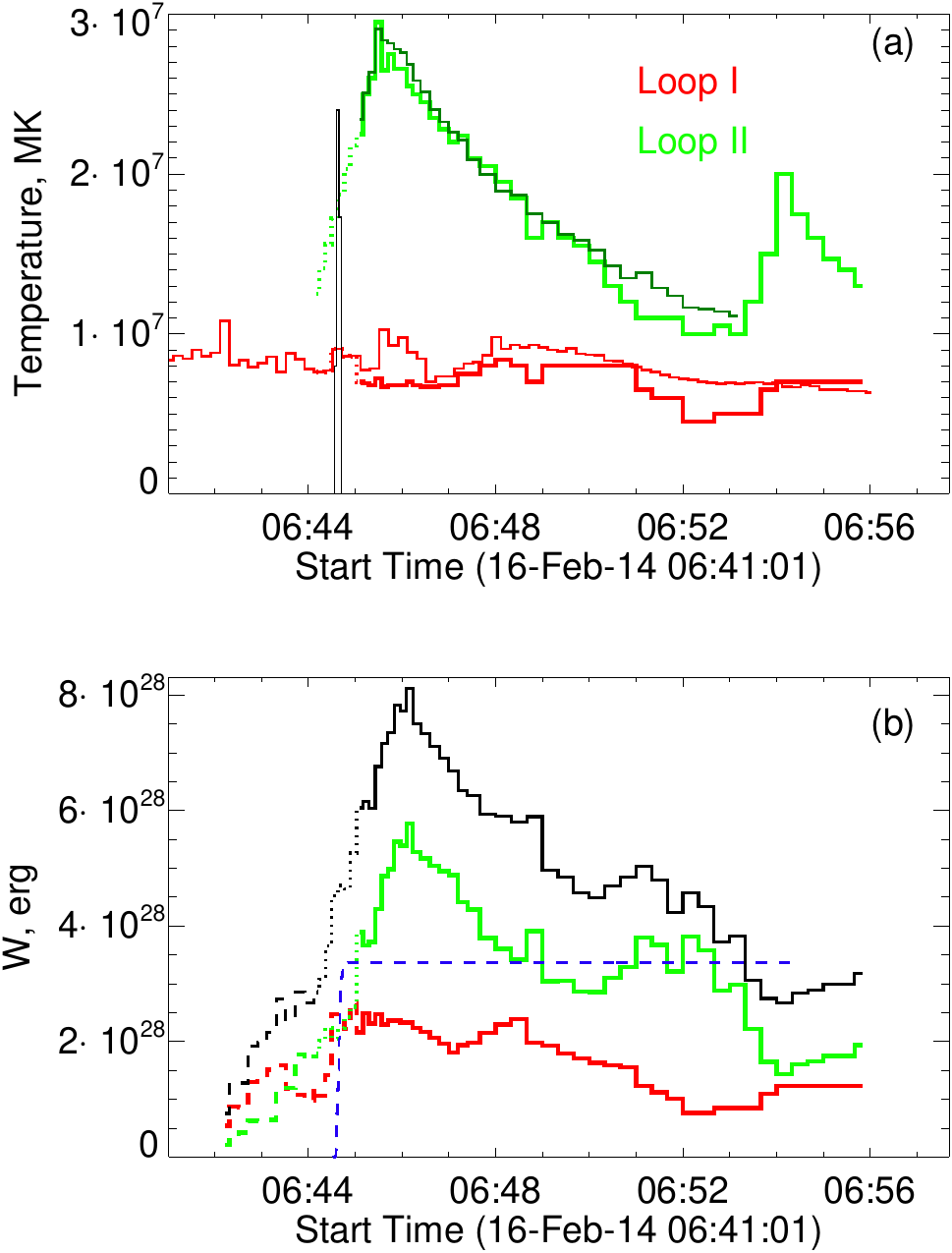}
\caption{
{(a) Evolution of the plasma temperature in Flux Tubes I (red) and II (green). The dotted green segment replicates the dotted segment in  Fig.\,\ref{f_temp_from_data}.  Dark green line shows the result obtained from \rhessi\  OSPEX 
fit (cf. Fig.\,\ref{f_temp_from_data}). Thin red line shows temperature obtained for Flux Tube I from AIA EM maps (cf. Fig.\,\ref{f_temp_from_data}). For the reference, the thin black line shows a conveniently scaled impulsive HXR emission  peak recorded by \kw . (b) Composite evolution of the thermal energy computed for Flux Tubes I (red) and II (green). The solid portions of the lines show the time range, when the model is constrained by both AIA and \rhessi\ data starting 06:45:06\,UT. Dashed segments show estimates of $W$ obtained using the number density derived from adjustment of the model to AIA EM maps and temperatures derived from the AIA temperature maps directly (shown in Fig.\,\ref{f_temp_from_data}). Green dotted segment uses the number density derived from adjustment of the model to the AIA EM maps and the temperature shown by the dotted green segment in panel (a).
The black line shows evolution of the total thermal energy obtained by adding up the contributions from both loops. The dashed blue line shows cumulative deposition of the nonthermal energy computed using the injection function determined in Section\,\ref{S_trapping}. }
\label{f_model_T_W}
}
\end{figure}

In the 3D model developed in \citet{2021ApJ...913...97F}, Flux Tube I is filled in with a cooler plasma, while Flux Tube II with a hotter plasma. Thus, the \rhessi\ derived $T_1$ and $T_2$ are associated with loops I and II, respectively. It is notable that the AIA DEM weighted temperature from Loop I matches $T_1$ reasonably well even though Loop I projects onto a leg of Loop II. On the contrary, $T_2$ is systematically larger than the DEM weighted temperature from Loop II by a factor of 2 or more. This is primarily due to dissimilar sensitivity of AIA and \rhessi\ to various temperature regimes: \rhessi\ is sensitive to the hottest plasma component, while AIA is progressively less sensitive to temperatures above 15\,MK. Thus, AIA DEM weighted temperature underestimates the true temperature from the hot plasma volumes. This means that the AIA derived DEM data mainly  constrain the EM in cooler Loop I, while \rhessi\ spectra constrain parameters of Loop II. 
Thus, to obtain evolving spatial distributions of the thermal plasma consistent with both AIA and \rhessi\ constraints, we employ the following workflow: 

\begin{enumerate}
    \item Define spatial distribution of the thermal number density in Flux Tube I (primarily) and II (auxiliary) to match DEM derived EM map at a given time frame.
    \item Repeat this step for another time frame until all time frames have been processed.
    \item For a given time frame, fix the thermal plasma density in Flux Tube I at the values found at the previous steps and determine {the} parameters of Flux Tube II and the temperature of Flux Tube I, such as to best match the  X-ray spectrum synthesized from the model with the observed one.
    \item Repeat this step for all time frames.
\end{enumerate}

\subsection{Thermal model description}
\label{S_therm_model_descr}

The standard description of the spatial distribution of the thermal number density $n$ in a given flux tube defined in the GX Simulator \citep{2023arXiv230100795N} is as follows:

\begin{equation}
    \label{Eq_n_def}
  n(x,y,s)=n_0 n_r(x,y)n_s(s),
\end{equation}
where
\begin{equation}
    \label{Eq_n_r_def}
n_r(x,y)= \exp(-(p_0 x/a(s))^2-(p_0 y/b(s))^2)
\end{equation}
is the transverse distribution of the thermal plasma (here the parameter $p_0$ can be different from a similar one for the nonthermal distribution defined in Section\,\ref{S_F_evolution}, while $a$ and $b(=a)$ are the same transverse reference sizes of the flux tube),
\begin{equation}
    \label{Eq_n_s_def}
n_s(s)= \exp(-(q_0((s-s_0)/l+q_1))^2)
\end{equation}
is the density distribution along the flux tube axis (again the free parameters $q_0$ and $q_1$ are generally different from those for the nonthermal electron distribution, while $s_0$ and $l$ are the same geometrical parameters of the magnetic flux tube). The free parameters {in Eqns. \ref{Eq_n_r_def} and \ref{Eq_n_s_def}} are also generally different for various flux tubes within the model. In contrast to the number density, the temperature of a loop is  set to a constant value, while different for various loops.

{The background corona, which surrounds the model flaring flux tubes, is adopted tenuous ($n\lesssim10^8$\,cm$^{-3}$) and relatively cool ($T=1$\,MK), such as its contribution to the synthesized emission or sampled plasma parameters is negligible. This is consistent with the procedure of back- and fore- ground subtraction described by \citet{2020ApJ...890...75M}.}

\subsection{Adjusting thermal model to data}

The thermal plasma distribution defined in the previous section, while a simplified analytical one, depends on many free parameters. As in the case with the nonthermal electron distribution, here we look for a solution with minimally possible number of varying parameters. To this end, we note that {the} \rhessi\ images do not show a significant evolution---the brightness peak remains at the same location, while the largest size of the source varies slightly back and forth. The AIA DEM-derived maps of the $EM$ display noticeable small-scale dynamics, but the overall location, size, and shape do not evolve too much \citep[see animated fig.\,7 in][]{2021ApJ...913...97F}. After several quantitative tests with the free parameters of the spatial distribution of the thermal number densities, we concluded that the parameters $p_0$, $q_0$, and $q_1$ can be kept constant in both flux tubes at the values: 
$p_0=1$, $q_0=1.7$, and  $q_1=0.3$ in Flux Tube I and 
$p_0=1$, $q_0=1.8$, and $q_1=0.1$ in Flux Tube II. Eventually, only the number densities $n$ and temperature $T$ were varied in the {tuning} process of the thermal structure.

\subsubsection{Adjusting  plasma density using AIA EM maps}\label{S_EM_AIA_tuning}

\gx\ offers several ways of rendering {the} 3D model volume, computing emissions, and performing quantitative  data-to-model comparison \citep{2023arXiv230100795N}. Perhaps, the most direct way is computing emission from the model volume and comparing this synthetic emission with the observed one. This approach does not work for the EUV emission from flaring loops, because the EUV emission is sensitive to details of {the} DEM distribution, while the thermal flare model is based on the description in terms of the number density and a single temperature in the given flux tube; see Section\,\ref{S_therm_model_descr}. For such cases \gx\ offers a direct sampling of the model volume, which provides line-of-sight (LOS) integrated number densities, $EM$, and $W$. Here we employ the synthetic $EM$ ``maps'' and compare them directly with the AIA DEM-derived $EM$ maps reported by \citet{2021ApJ...913...97F}. In many time frames the observational $EM$ maps are dominated by contribution from Flux Tube I. In such cases, we vary the thermal number density in that loop such as to minimize the residuals and normalized residuals between the synthetic and observational map, while the number density in the second loop remains unconstrained (only an upper limit can be determined to avoid any excessive contribution from it). In other time frames {the} contribution from Flux Tube II is not negligible and can be constrained by maximizing the cross-correlation coefficient between the synthetic and observational maps. 
In the process of the thermal model adjustment we considered the EM map area above a threshold of 12\% of the peak value, both model and observed, although in many time frames the data-to-model agreement holds to much lower threshold. Evolution of the thermal densities obtained this way is shown in Figure\,\ref{f_model_n_EM}a {as} thin dark red and dark green lines.

\subsubsection{Adjusting  thermal model to the \rhessi\ spectral data}\label{S_rhessi_tuning}

At this stage we {vary thermal plasma parameters such as to match the synthetic, FOV-integrated X-ray spectrum to the \rhessi\ spectral} data to better constrain the number density and quantify the temperature in Flux Tube II, because it contains the hottest plasma in this flare to which \rhessi\ is most sensitive. For Flux Tube I we adopted the thermal number density to be identical to that constrained by AIA EM maps in Section\,\ref{S_EM_AIA_tuning}, with the only adjustment in the time stamps: to each \rhessi\ time stamp we assigned the density from the closest AIA time stamps because they do not coincide with each other. The result is shown in Figure\,\ref{f_model_n_EM}a by the red thick line. Having the density in Flux Tube I fixed, we adjusted the density and temperature in Flux Tube II to minimize the residual in the high-energy part of the X-ray spectrum (above 7\,keV) and then adjusted the temperature in Flux Tube I such as to minimize the residual in the low-energy part of the  X-ray spectrum (3--6\,keV).

{The thick green line in } Figure\,\ref{f_model_n_EM}a displays the evolution of the thermal number density in Flux Tube II. This evolution is consistent with the one estimated using the AIA EM maps (dark green line), but reveals a lot more details in the light curve. Figure\,\ref{f_model_n_EM}b shows the EM of Flux Tube I, constrained by the AIA data, and Flux Tube II, constrained by {the} \rhessi\ data.

Figure\,\ref{f_model_T_W}a displays {the} evolution of the temperatures in the flux tubes by thick red and green lines. For the reference, thinner red and dark green lines show the corresponding temperatures derived from the AIA EM and \rhessi\ data (shown earlier in Figure\,\ref{f_temp_from_data}). The model-derived and data-derived temperatures show a reasonable agreement between each other.

\begin{figure}\centering
\includegraphics[width=1\linewidth]{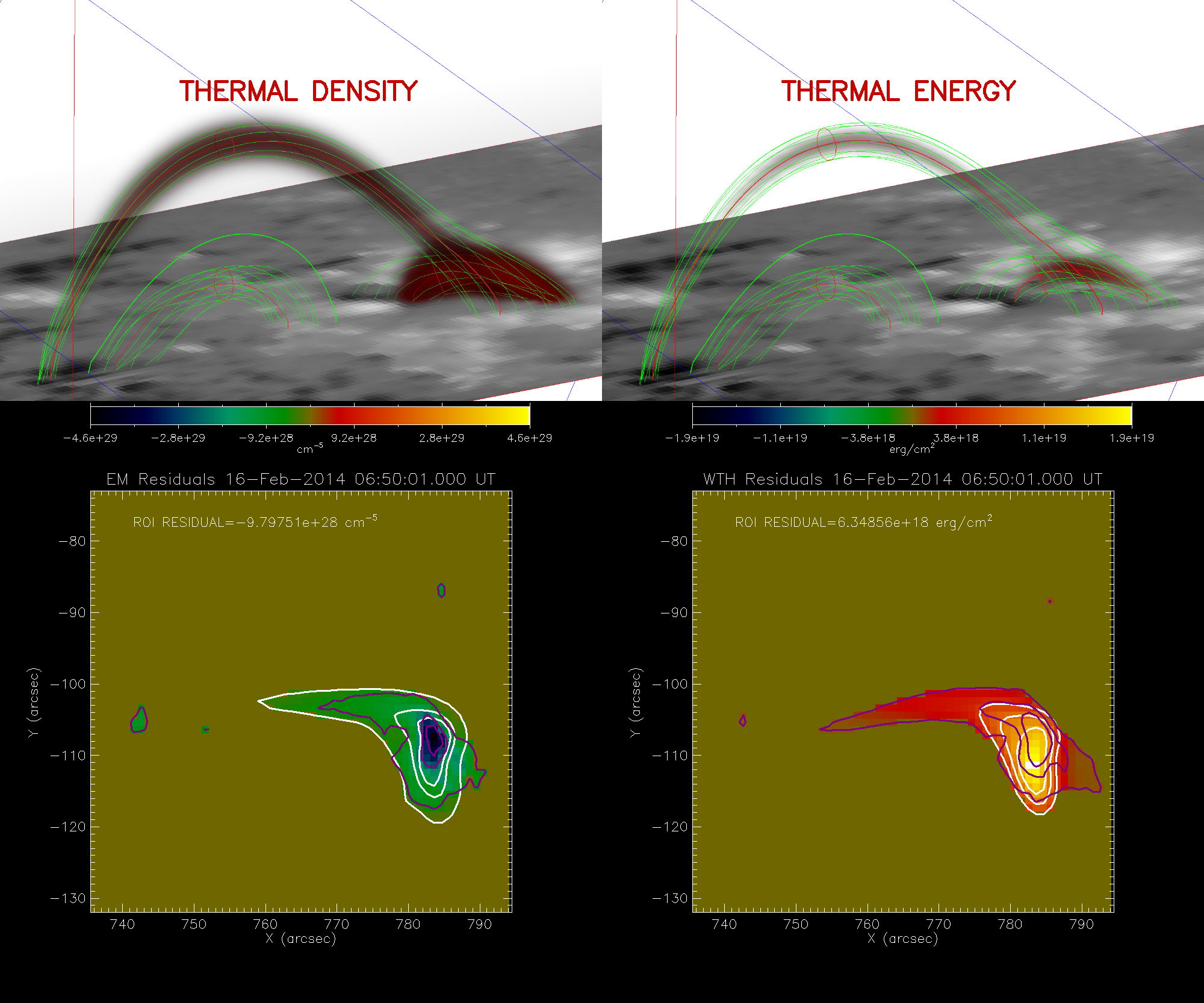}
\caption{
{Perspective view of the 3D distribution of the thermal density (top left) and thermal thermal energy (top right). The bottom panels show the residuals between the LOS-integrated model (white contours) and AIA-derived (violet contours) thermal plasma density (left) and thermal energy (right). The contours represent the 30\%, 50\%, and 70\% levels. {An animation is included with this figure. The video shows a full sequence of the time frames with the same layout of the panels as in the front figure. The video
duration is 15 s.} }
\label{f_3D_dens_W}
}
\end{figure}

Figure\,\ref{f_model_T_W}b displays {the} evolution of the thermal energy of the flare. The solid segments of the curves show the model results constrained by both AIA and \rhessi\ data starting {at} 06:45:06\,UT. At earlier time, the model is only constrained by the AIA data. The dashed segments show the curves computed based on the model AIA-constrained density and DEM-weighted temperatures shown by  red and green lines in Figure\,\ref{f_temp_from_data}.
The dotted segments employ the Flux Tube II temperature extrapolated from the the \rhessi\ spectral fit, shown by the dotted green line in panel (a). The black curve represents the total thermal energy obtained by adding up contributions from both flux tubes. For comparison, the cumulative deposition of the nonthermal energy is shown by the dashed blue line. This deposition is computed by integration of the injection function $G$ over the volume, energy, and eventually time. 
Figure\,\ref{f_3D_dens_W} offers a 3D visualization of the thermal density and energy evolution in the the model flux tubes.

\section{Discussion}
\label{S_discussion}

This study presents the first data-constrained 3D modeling of a solar flare evolution {that} quantitatively addresses electron acceleration, their  transport, and heating of the ambient plasma. We also evaluate the energy budget and partitions between various energy components and different flaring loops. We compare the energy budget obtained from the (2D) data analysis and the one obtained here from the data-validated evolving 3D model.

\subsection{Acceleration of electrons}
\label{S_disc_accel}

Our analysis reveals several interesting features of the accelerated particle population. Their number density appears rather high, $n_b\approx10^8$\,cm$^{-3}$ at its spatial peak, starting from the very first time frame when the nonthermal \mw\ emission becomes detectable. At this point, the spatial distribution of the nonthermal electrons is very compact (see Fig.\,\ref{f_nonth_el_evol}) and appears at a position where the two flaring loops intersect (in projection); thus, the energy release is likely due to interaction between these two flaring loops. The peak number density remains roughly constant during early raise phase of the radio burst, while the observed increase of the radio flux is driven by {the} increase of the volume occupied by the nonthermal electrons. This spreading of the nonthermal electrons over the flaring volume begins in a more compact, dense loop (Flux Tube I) and then continues in a more extended and more tenuous loop (Flux Tube II). In Flux Tube II, both number density and the occupied volume raise during the burst raise phase. The nonthermal number density reaches high values up to $n_b\approx1.2\times10^{10}$\,cm$^{-3}$ at its spatial peak in Flux Tube II, which is comparable with the thermal number density in this loop.

The {spatial} spread of the nonthermal electrons cloud in space can be due either to their spatial diffusion \citep[cf.][]{2018ApJ...859...17F} or the spread of the magnetic reconnection/energy release itself. In the case {analyzed here}, it is unlikely that the spatial diffusion plays a noticeable role, because this would imply a long trapping time, which we proved not {to be} the case. If the reconnection spreading plays a role, then this spreading has to occur with roughly the Alfven speed. Given that our 3D model provides the magnetic field and the number density everywhere in the volume, we can straightforwardly compute  the Alfven velocity everywhere. It appears {being} about $3\times10^8$\,cm\,s$^{-1}$ in both flux tubes, which is indeed consistent with the identified speed of the electron cloud raise. 

Let us discuss why the acceleration efficiency $\eta = n_b/(n+n_b)$ is strongly different in the two flaring loops: $\eta(I)=1.82\times10^{-3}$, while $\eta(II)\sim1$. We do not have any reliable means to nail down the microscopic acceleration mechanism in this event; thus, we consider the processes of the energy gain and loss at a very basic level. We note that, regardless of the acceleration mechanism, the only available accelerating force is the electric force, which has to do a work on a charged particle over some distance to accelerate it to a certain energy. To gain this acceleration, the particle loss must be less than the gain over this distance. The balance between the accelerating electric force and the loss due to Coulomb collisions is described in terms of the Dreicer field \citep{PhysRev.115.238}. If the electric field is less than the Dreicer one, then only a fraction $\eta_{run}$ of electrons from the Maxwellian distribution tail 

\begin{equation}
 \eta_{run}\propto\exp(-v_{run}^2/2v_T^2), \quad v_{run}=v_T\sqrt{(E_D/E)  },
\end{equation}
can be accelerated. 

If the electric field is larger than the Dreicer one, then bulk acceleration of literally all electrons can occur. The Dreicer field is defined by the plasma temperature and density \citep[e.g., Eq.(11.1) in][]{Fl_Topt_2013_CED}, which are known everywhere in our 3D model. Thus, we can compute the Dreicer fields in the two flaring loops to be $E_D(I)\approx 4\times10^{-4}$\,V\,cm$^{-1}$ for $n=5.5\times10^{10}$\,cm$^{-3}$; $T=8$\,MK and $E_D(II)\approx 3\times10^{-5}$\,V\,cm$^{-1}$ for $n\sim1\times10^{10}$\,cm$^{-3}$; $T\sim20$\,MK. To accelerate a fraction of $\eta(I)=1.82\times10^{-3}$ as a run away population from the Maxwellian tail $\eta=\exp(-v_{run}^2/2v_T^2)\approx\exp(-6.3)=1.82\times10^{-3}$, we need  {an} electric field of about $E_D(I)/12.6\approx 3.27\times10^{-5}$\,V\,cm$^{-1}$, which is comparable with $E_D(II)$ computed above. Thus, for the same electric field, a bulk run away acceleration is likely in Flux Tube II. We conclude that, if the magnetic reconnection due to interaction between these two flaring loops results in comparable electric fields in both loops, the run away electron fractions will be much different in these flux tubes, in quantitative agreement with the acceleration efficiencies identified here. These properties of the particle acceleration offer additional constraints to numerical acceleration models \citep{2020PhPl...27j0601D,PhysRevLett.126.135101,2022ApJ...932...94D}, which have to explicitly take into consideration the energy loss due to Coulomb collisions.


\subsection{Electron trapping}

Our analysis of the nonthermal electron transport in Flux Tube II that dominates the impulsive \mw\ emission, indicates that no significant trapping takes place there. The evolution of the nonthermal electron population does not show any noticeable spectral evolution other than an increase of the maximum electron energy very early in the burst. A trapping model with the constant escape time ($\tau=1$\,s) and constant spectral index ($\delta=3.9$) is quantitatively consistent with the \mw\ data. 

Flux Tube II has a mirror ratio about $m=5$; thus, isotropic injection within a weak diffusion regime (weak angular scattering) would result in a rather efficient trapping, which is not observed. The strong diffusion regime would result in a slow diffusive propagation of the electrons in the loop, which implies a rather long escape time too. Therefore, the observed short escape time would require either anisotropic injection along the magnetic field and the free-streaming escape to the foot points with the escape time of the order of $L/v$, or the moderate diffusion regime with the escape time $mL/v$, where $L\approx3\times10^9$\,cm is the half length of the loop and $v$ is the speed of the nonthermal electron. To reconcile these escape times with the observed value of 1\,s, the free streaming regime would imply electrons with velocity about $3\times10^9$\,cm\,s$^{-1}$ ($\sim$3\,keV), while the moderate diffusion regime---$1.5\times10^{10}$\,cm\,s$^{-1}$ ($\sim$80\,keV). The moderate diffusion regime looks more plausible based on this energy consideration.

\subsection{Plasma heating and cooling}

Figure\,\ref{f_model_T_W}b shows {the} evolution of the thermal energies in Flux Tubes I and II, computed by the volume integration of the 3D distributions of our data-validated model. This plot also shows the total thermal energy in these loops and the  energy deposition by the injected nonthermal electrons described by the injection function $G$ determined in Section\,\ref{S_trapping}. Several important conclusions about plasma heating can be made: (1) the amount of the deposited nonthermal energy is insufficient to drive the plasma heating; (2) timing of the heating also requires additional energy sources as the heating starts earlier and lasts longer than the impulsive injection of the nonthermal electrons; (3) the nonthermal energy deposition is, however, sufficient to drive the plasma heating during the impulsive peak. These findings are consistent with those reported by 
\citet{2021ApJ...913...97F}. We do not have any reliable means to identify {the} mechanisms of this additional plasma heating---this can be a ``direct'' heating due to magnetic reconnection, or energy deposition due to accelerated ions, or something else.  Note that the direct heating is often attributed to a superhot coronal component \citep{2010ApJ...725L.161C,2014ApJ...781...43C}, while here we  observe numerous direct heating episodes resulting in rather modest plasma temperature around 10\,MK.

Cooling of a hot flaring plasma is typically dominated by the heat conduction \citep[e.g.,][]{Aschw_2005}. \citet{2016ApJ...822...71F} presented the classical (Spitzer) conductive cooling time described by Eq. (4.3.10) from \citet{Aschw_2005} in a convenient form:

\begin{equation}
\label{Eq_conduct_decay_time}
   \tau\simeq2.4\cdot10^3~{\rm [s]}
    \left(\frac{L}{10^{10}~{\rm cm}}\right)^{2}\left(\frac{n_e}{10^{10}~{\rm cm}^{-3}}\right)
   \left(\frac{10^7~{\rm K}}{T}\right)^{5/2},
\end{equation}
from which we can straightforwardly estimate $\tau\sim350$\,s for FT I and  $\tau\sim60$\,s  for FT II. Figure\,\ref{f_model_T_W}a shows that the Flux Tube II cooling occurs a factor of four slower than this estimate, while Flux Tube I does not show any noticeable cooling at all---here the decrease of the thermal energy is mainly associated with decrease of the plasma density rather than temperature. 

It  is known  that the Spitzer conduction can be suppressed if the conduction occurs in a ``locally limited'' or ``free streaming'' regimes. 
According to fig. 6 from \citet{2009A&A...498..891B}, {the} parameters of both loops correspond to the ``locally limited'' heat conduction, when the heat flux is reduced by a certain fraction $\varrho(\cal R)$, where ${\cal R}=\lambda/L$, $\lambda=5.21\times10^3 T^2/n$ is the mean free path of the thermal electrons, and $L$ is the temperature scale length, adopted here to be a half of the loop length. The cooling time is enhanced by the same factor $\varrho(\cal R)$. Using Eq.\,(4) from \citet{2009A&A...498..891B}, we computed the ranges of the correction factor $\varrho_1$ and $\varrho_2$ for our flux tubes over the course of the plasma evolution and found that always $\varrho_1>0.92$ and $\varrho_2>0.8$. Thus, the effect of the locally limited heat conduction cannot account for the observed slowing the plasma cooling down. One possibility to have such an extended cooling phase is a sustained energy deposition to the flaring plasma during the cooling phase. Indeed, the cooling of both loops is not monotonic; there are episodes of   energy increase, which require an energy input.

\subsection{Energy partitions.  Model vs data analysis.}

Here we discuss the evolving energy partitions between (i) nonthermal and thermal components and (ii) Flux Tubes I and II. The nonthermal energy is only detected during the flare impulsive phase, which lasted about half a minute. During this impulsive phase, a total of about $W_{nth}\approx3.37\times10^{28}$\,erg of nonthermal energy was released with a peak energy deposition rate of about $\dot{W}_{nth}\approx6.13\times10^{27}$\,erg\,s$^{-1}$. These numbers are consistent with those estimated by \citet{2021ApJ...913...97F} using a simplified approach ($\dot{W}_{nth}\approx5.87\times10^{27}$\,erg\,s$^{-1}$ and  $W_{nth}\approx4.7\times10^{28}$\,erg). The total released nonthermal energy is less than the peak thermal energy ($W_{th}\approx8.11\times10^{28}$\,erg) by more than a factor of two. The thermal energy increase during/right after the impulsive phase roughly coincides with the amount of released nonthermal energy. Thus, the amount of thermal energy due to direct heating (or other mechanism(s) not related to the nonthermal electron loss) is about $W_{th, dir}\approx4.74\times10^{28}$\,erg, which is about 40\% larger than the total deposited nonthermal energy. The fact that the nonthermal energy deposition is insufficient to heat the flaring plasma is independently confirmed by {the} evolution of the thermal energy: it starts to increase well before the nonthermal impulsive phase, reaches the peak about 90\,s after the impulsive peak, and shows several episodes of the thermal energy increase after that. In particular, the thermal energy (and the temperature) increases after $\sim$06:54\,UT, when {the} AIA parameter maps show appearance of a new connectivity loop in {the} late flare decay phase; perhaps, due to one more reconnection episode. The thermal energy evolution is driven by a trade-off between sustained but varying plasma heating and permanently acting energy loss. This means that the total amount of ``directly'' released thermal energy is larger than the estimated value of $W_{th, dir}\approx4.74\times10^{28}$\,erg.

The released energy divided mainly between two flare loops (I and II), while a third one identified in \citet{2021ApJ...913...97F} played a more minor role and is not discussed here. The thermal energy is divided between these two loops in comparable amounts: $W_{I th, \max}\approx2.49\times10^{28}$\,erg and $W_{II th, \max}\approx5.78\times10^{28}$\,erg. The difference between these two values, $\approx3.45\times10^{28}$\,erg, is very close to the released nonthermal energy $W_{nth}\approx3.37\times10^{28}$\,erg, which was almost entirely released in Flux Tube II. Thus, we conclude that the thermal energy due to ``direct'' heating is almost equally partitioned between Flux Tubes I and II. In contrast, as it has been shown, the nonthermal energy divided highly unevenly between these two loops. The reason for this imbalance might be associated with different values of the thermal number density and temperature in these two loops as discussed in Section\,\ref{S_disc_accel}: if the acting electric field (available for electron acceleration) is comparable in both loops, while they have dissimilar Dreicer fields, the accelerated electron fractions can be strongly different, as observed. Thus, only a minor fraction of the available thermal electrons in Flux Tube I is accelerated to nonthermal energies, while most of released energy goes to the plasma heating. In contrast, a large fraction of available electrons is accelerated in Flux Tube II during the impulsive phase (presumably, when the most prominent energy release takes place), while the direct plasma heating dominates here before and after the impulsive phase (presumably, because the power of the free energy source is smaller than during the impulsive phase).


In conclusion, we emphasize {the} advantages of studying these energy partitions based on the 3D model in comparison with the study based on the 2D data analysis performed in \citet{2021ApJ...913...97F}. Firstly, the flux tubes project one onto the other. Thus, {an} analysis of {the} 2D maps does not permit spatial separation of contributions from these two loops. And secondly, the 3D model validated by comparison with all available observational constraints is free from the LOS ambiguity; thus, permitting us to investigate the plasma properties including the energy partitions in the 3D realm.


\acknowledgements
This work was partly supported 
by NSF grants AGS-2121632,  
{1743321 }
and AST-2206424,  
and NASA grants
80NSSC20K0627, 
80NSSC19K0068, 
and 80NSSC23K0090, 
to New Jersey Institute of Technology. 
GM (\sdo/AIA and \rhessi\ data analysis) was supported by RSF grant 20-72-10158.

\newpage

\bibliography{flares_references,2017sep10,fleishman}

\end{document}